\documentclass{aa}
\usepackage{psfig,epsf,amssymb,amsmath}
\begin{document}

\title{Origin of non-keplerian motions of masers in NGC 1068}
\titlerunning{Origin of non-keplerian motions of masers in NGC 1068}
\author{Jean-Marc Hur\'e\inst{1,}\inst{2,}
}
\offprints{Jean-Marc.Hure@obspm.fr}
\institute{LUTh (FRE 2462 CNRS), Observatoire de Paris-Meudon, Place Jules Janssen, F-92195 Meudon Cedex
\and
Universit\'e Paris 7 Denis Diderot, 2 Place Jussieu, F-75251 Paris Cedex 05}
\date{Received ???; accepted ???}

\abstract{
We demonstrate that the ``sub-keplerian'' rotation curve of maser spots in NGC 1068 can be explained by the gravitational attraction of the disc orbiting the central black hole. Possible parameters matching observations are: black hole mass $\simeq  (1.2 \pm 0.1) \times 10^7$ M$_\odot$, disc outer edge $\gtrsim 1.3$ pc, aspect ratio $3 \times 10^{-3} \lesssim \frac{H}{R} \lesssim 0.3$, surface density $\Sigma \propto R^{-1.05 \pm 0.10}$, and disc mass $\simeq (9.4 \pm 1.6) \times 10^6$ M$_\odot$. The physical conditions required for the excitation of masers are fulfilled,
and the outer disc would stand in a gravitationally marginally stable state.
}

\maketitle

\section{Introduction}

Even though direct images are missing, evidences that Active Galactic Nuclei (AGN) host a super-massive black hole surrounded by a parsec-size accretion disc are compelling. In a few objects, the spatial resolution is just sufficient to unveil the outermost regions of the central engine at the sub-parsec scale thanks to the detection of water maser emission (Greenhill, 2002). The best example is NGC 4258 where masers have been associated with a thin disc in quasi-perfect keplerian rotation (Miyoshi al. 1995). In contrast, the azimuthal velocity of masers in NGC 1068 varies with the radius $R$ as $R^{-0.31\pm 0.02}$  according to Greenhill et al. (1996) who mentioned the possible role of the disc/torus self-gravity.
 To our knowledge, this hypothesis has never been explored, even qualitatively. We show in section \ref{sec:basiscideas} that a sub-keplerian velocity exponent is theoretically expected close to the outer edge of a (low mass) disc when considering self-gravity. In section \ref{sec:ngc1068}, we report possible parameters for the disc and for the black hole in NGC 1068 such that the computed velocity match the observed motion of maser spots. A few remarks and possible improvements are found in the last section.

\section{Basic ideas}
\label{sec:basiscideas}
\subsection{Gravity, orbiting velocity and velocity exponent}
\begin{figure}
\psfig{figure=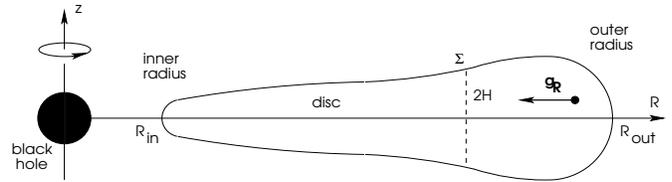,width=8.7cm}
\caption{Schematic view (not to scale) of the system.}
\label{fig:img}
\end{figure}

We consider the steady state system\footnote{In the following, quantities labeled with ``bh'' refer to the black hole, and those labeled with ``disc'' refer to the disc.} schematically pictured in Fig. \ref{fig:img}, made of a black hole with mass $M_{\rm bh}$ and Schwarzschild radius $R_{\rm S}=\frac{2GM_{\rm bh}}{c^2}$ surrounded by an axi-symmetrical, finite size gaseous disc with inner edge $R_{\rm in}$, outer edge $R_{\rm out}$, semi-thickness $H$, surface density $\Sigma$ and mass $M_{\rm disc}=qM_{\rm bh}$. Assuming that gravity balances the centrifugal force in the disc, then the rotation velocity $v_\phi$ of gas at cylindrical distance $R$ from the center obeys the Euler equation
\begin{equation}
v_\phi^2 =  - Rg_R \qquad (> 0),
\label{eq:vphi}
\end{equation}
where $g_R=g^{\rm bh}_R+g_R^{\rm disc}$ is the radial component of the total (black hole + disc) gravitational acceleration. Note that we neglect the motion of the system around its center of mass (although possibly significant) and consider the black hole as ``fixed''. Further, the mass distribution in the disc is such that $g_R < 0$ at the outer edge.
We can define the total ``velocity exponent'' $k$ as
\begin{equation}
k \equiv \frac{\partial \ln v_\phi}{\partial \ln R} = \frac{1}{2}\left[1+ \frac{\partial \ln (-g_R)}{\partial \ln R}\right],
\label{eq:alpha}
\end{equation}
so that the azimuthal velocity varies as $R^k$, locally. In particular, for a black hole only, one finds the keplerian value $k^{\rm bh} = 1-\frac{3R^2}{2r^2}$ where $r=\sqrt{R^2+z^2}$ is the spherical radius (and $k^{\rm bh}=-\frac{1}{2}$ at the equatorial plane $z=0$).

\subsection{Combining a black hole and a disc: edge effects}
\label{subsec:comb}

There is no reliable, regular expression\footnote{It is worth noting that any expression for $g^{\rm disc}_R$  established assuming a zero-thickness disc inevitably diverges at the edges (Mestel, 1963) and must not be used in the disc inside and in the neighborhood as well. This is why we shall employ here a numerical tri-dimensional approach which provides everywhere with a regular self-gravity field. \label{note:3d}} for the radial field $g_R^{\rm disc}$ due to a disc (as a tri-dimensional object) with given shape and density distribution, and thus no straightforward expression of the disc velocity exponent $k^{\rm disc}$ is available. For the present system however, the relation
\begin{equation}
k - \frac{1}{2} =\left[k^{\rm bh} - \frac{1}{2} \right]\frac{g^{\rm bh}_R}{g_R}+\left[k^{\rm disc} - \frac{1}{2} \right]\frac{g^{\rm disc}_R}{g_R} 
\label{eq:mixture}
\end{equation}
can easily be established from Eq.(\ref{eq:alpha}). We thus see that a departure to the keplerian motion (that is $k \ne k^{\rm bh}$) can arise in two opposite situations: i) when the disc self-gravity exceeds the gravity due to the black hole, or ii) when the disc self-gravity is much smaller than the central gravity, but $|k^{\rm disc}| \gg 1$. In the extreme, the first case suggests that a non keplerian rotation can be produced by a disc only (without a black hole; i.e. $q \rightarrow \infty$). Here, we shall adopt the second point of view  which rather corresponds to configurations with a mass ratio $q \lesssim 1$.

\begin{figure}
\psfig{figure=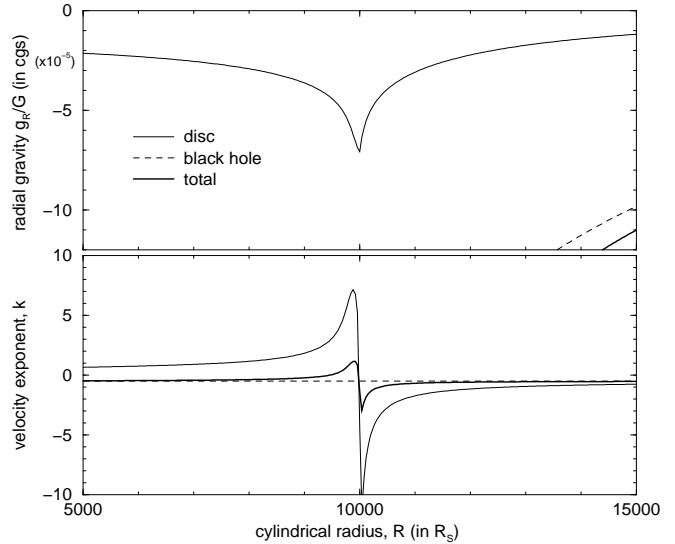,width=8.7cm}
\caption{Example of radial field $g_R$ ({\it top}) and velocity exponent $k$ ({\it bottom}) at the equatorial plane due to a $10^8$ M$_\odot$ black hole and to a low mass standard-type disc with outer edge located at $10^4 R_{\rm S}$ (see text for more details).}
\label{fig:typical}
\end{figure}

The point is that, depending on the mass distribution in a disc, large values of $k^{\rm disc}$ can occur at the edges, and especially at the outer edge of interest here, even when $q \ll 1$. The physical reason is that radial gravity is generally vanishingly small well in the middle of a disc because the gravitational contributions of matter located interior and exterior to any point essentially cancel out (unless very steep density gradients). Conversely, near an edge, there creates an imbalance between these two contributions, and the resulting field can have a quite large amplitude. Hence a gradient of radial self-gravity. An illustration of this "edge effect" is given in Fig. \ref{fig:typical} which displays the radial field and velocity exponent in a system with mass ratio $q=0.1$. In this example, we have adopted parameters typical for an AGN standard disc (e.g. Frank, King \& Raine, 1992), namely black hole mass $M_{\rm bh}=10^8$ M$_\odot$, disc inner edge $R_{\rm in}=3 R_{\rm S}$ and outer edge $R_{\rm out}=10^4 R_{\rm S}$, aspect ratio $\frac{H}{R}=0.01 \times f(R)$ and surface density $\Sigma \propto R^{-3/10} \times f(R)$, where we have introduced the $f$-function in order to make the disc thickness and density gradually decrease to zero over a disc scale height typically (i.e. $f \simeq 1$ over almost all the disc extent and $f=0$ outside $[R_{\rm in}, R_{\rm out}]$). This is necessary to avoid a sharp discontinuity between the disc and the ambient medium which would not be realistic at all. Further, we impose that the vertical stratification of matter between the disc mid-plane and its surface is quadratic with the altitude. This conveniently mimics the Gaussian profile expected in a vertically isothermal disc as well as the solution of the plane Lane-Emden equation for a vertically self-gravitating disc (e.g. Iba\~nez \& Sigalotti, 1984). Quite importantly, we point out that both the adopted prescription for $f(R)$ and the assumption regarding the vertical stratification have no noticeable influences on the global results within physically acceptable limits, so that {\it issues raised here are not artefacts due to discontinuity effects}. Finally, the radial gravitational acceleration due to the disc is computed numerically using the accurate Poisson 3D-solver described in Hur\'e (2002). 

We see that, although the disc is ``light'' with respect to the black hole and central gravity fully dominates, $k$ deviate markedly from the keplerian value near the outer edge. A general property (met here) is that the rotation is characterized by a ``sub-keplerian velocity exponent'' for $R \lesssim R_{\rm out}$ in the sense that $k \ge -\frac{1}{2}$, whereas this is just the opposite for $R \gtrsim R_{\rm out}$.

\begin{figure*}
\psfig{figure=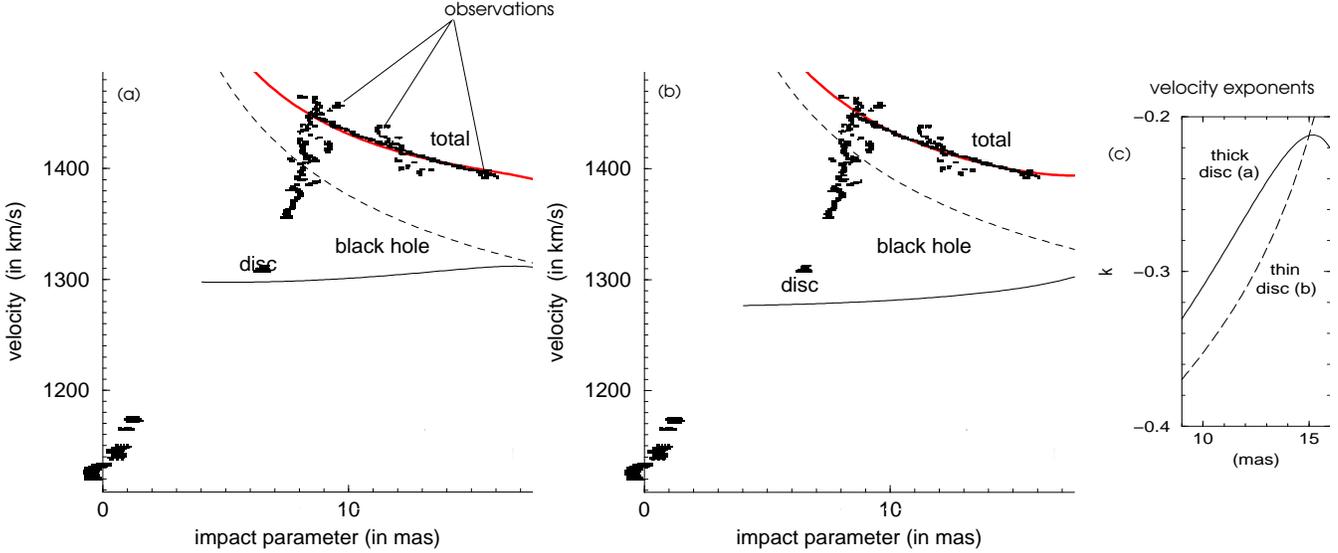,width=17.7cm}
\caption{Theoretical rotation curve ({\it bold line}) expected at the surface of the disc in NGC 1068 compared to observations (adapted from Greenhill et al. 1996): ({\it a}) thick disc solution, ({\it b}) thin disc solution, and ({\it c}) associated velocity exponents (see also Tab. \ref{tab:cmp}). Also shown are the contributions of the disc ({\it plain line}) and of the black hole ({\it dashed line}).}
\label{fig:cmp}
\end{figure*}

\section{The case of NGC 1068}
\label{sec:ngc1068}

\subsection{Origin for the non-keplerian motions}
\label{subsec:nkm}

NGC 1068 is probably one of the best studied\footnote{See for instance the proceedings of the workshop published in the '97 special issue of {\it Astrophysics and Space Science}.} among type-2 Seyfert galaxies. In this object with systemic velocity of $\sim 1126$ km/s and distance of $15$ Mpc, water maser emission has been detected at the parsec scale. Some emission features would originate from the surface of a disc, along a diameter perpendicular to the line of sight, in the range $0.65 - 1.1$ pc (Greenhill et al. 1996). This corresponds to a masing disc extending from $\sim 9$ to $15$ milli-arcseconds (mas) in the sky. From the rotation curve of water maser spots, the dynamical mass enclosed within $1.1$ pc is estimated to $\sim 1.5 \times 10^7$ M$_\odot$ and the best fit for the velocity exponent yields $k_{\rm obs.} = -0.31 \pm 0.02$.

The significant departure to Kepler law means that this dynamical mass may not be that of the black hole. Greenhill et al. (1996) suggested that disc self-gravity could be responsible for this non-keplerian motion. Two other, plausible explanations have be defended: radiation pressure effects (Pier \& Krolik, 1992) and extra gravitational attraction by a stellar cluster (Kumar, 1999). However, the disc self-gravity hypothesis is quite natural, at least because NGC 1068, as an active galaxy, hosts a black hole {\it and a disc with a certain mass}. It is also corroborated by the arguments exposed in \S \ref{subsec:comb}. From this point of view, the condition $k_{\rm obs.} > k^{\rm bh}$ (i.e. a sub-keplerian exponent) implies that the maser spots in this AGN rotate inside the disc (see Fig. \ref{fig:typical}), and not outside (which would mean a "super-keplerian" exponent). We can thus conclude that the outer edge of the masing disc is larger than $\sim 1.1$ pc.

\subsection{``Inverting'' the rotation curve}

Matching observational data means that one must find the right magnitude of the orbiting velocity, and not only the velocity exponent. As $v_\phi$ depends on the distribution of matter in space (see Eq. (\ref{eq:vphi})), we see that the problem is to determine, if possible, which mass distribution(s) can cause the observed motions | a common problem in Astrophysics. More formally, this means the inversion of some relation
\begin{equation}
v_\phi \equiv v_\phi(R,z;M_{\rm bh}, R_{\rm in}, R_{\rm out}, \Sigma, H, \dots),
\label{eq:f}
\end{equation}
which obviously involves several parameters. In such a problem, not a single solution does exist but many, and finding all solutions would require the exploration of the whole parameter space. This task is complex and well out of the scope of this paper. However, preliminary investigations have revealed that i) the location of the inner edge has no influence provided $R_{\rm in} \ll R_{\rm out}$, and it is set to $3 R_{\rm S}$, ii) the disc thickness is not a critical parameter, provided the disc is geometrically thin (namely $\frac{H^2}{R^2} \ll 1$), and iii) the outer edge should stand beyond $1.1$ pc (see \S \ref{subsec:nkm}).

To simplify more the quest of solution, we further assume an uniform aspect ratio $\frac{H}{R}$, with $0.3$ as a firm upper limit (Gallimore et al., 1996), and we impose that the surface density is of the form $\Sigma = \Sigma_0 R^s$ (we keep on using the smoothing $f$-function; see \S \ref{subsec:comb}). Besides, we set $z=\pm H$, since we are interested in the velocity field at the disc surface where masers are assumed to be excited.
 Then, remains to be constrained mainly the central mass, the surface density and the location of the outer edge.

\begin{table}
\begin{tabular}{lcc}
parameters & thick disc & thin disc \\ \hline
$M_{\rm bh} (10^7 $M$_\odot$)  &   $1.2 \pm 0.1$       &  $1.2 \pm 0.1$    \\
$R_{\rm out}$ (pc) &  $\gtrsim 1.3$ ($18$ mas) & $\gtrsim 1.5$ ($21$ mas)\\
$H/R$ &  $0.3$ &  $\ll 0.3$ \\
$\Sigma$ ($0.65$ pc) (g/cm$^2$) & $\sim 470$ & $\sim 285$ \\
$n$ ($0.65$ pc) (cm$^{-3}$)  &  $\sim 10^8$ & $\gg 10^8$\\
$s$                      &  $-1.05 \pm 0.1$ &  $-1.00 \pm 0.05$ \\
$q$ ($0.65$ pc)    & $0.49$ &  $0.28$ \\
$q$ ($R_{\rm out}$)  & $0.91$ & $0.65$ \\
$k$                       & $-0.28 \pm 0.06$ &  $-0.29 \pm 0.08$   \\ \hline
\end{tabular}
\caption{Parameters inferred for two ``extreme'' solutions matching observations (see also Fig.\ref{fig:cmp}).}
\label{tab:cmp}
\end{table}

\subsection{Properties of the outer disc and central mass}

Two extreme solutions will be discussed here: a thick disc solution with the maximum value for aspect ratio, and a thin disc solution. Table \ref{tab:cmp} lists the parameters deduced for the disc and for the black hole. As Fig. \ref{fig:cmp} shows, both models fit very well observations, despite the total velocity index $k$ is far from being constant (but is of order of $-0.3$ on average). It turns out that i) a single value for the black hole mass is required, ii) $\Sigma$ decreases roughly as $R^{-1}$, iii) the disc outer edge is larger than $1.3$ pc (or $\sim 18$ mas), and iv) the disc mass is close the black hole mass. Note that the data reported here concern the outer disc and must not extrapolated down to the central black hole, but only over one or two decade(s) in radius.

These results deserve two comments. First, the physical conditions required for the emission of water maser seem met (Neufeld, Maloney \& Conger, 1994). Actually, at $0.65$ pc, adopting $\Sigma \sim 380$ g/cm$^2$ as a mean value (see Tab. \ref{tab:cmp}), the typical number density is
\begin{equation}
\left<n\right> \sim \frac{\Sigma}{2 H \mu m_{\rm H}} \simeq \frac{3 \times 10^7}{\left(\frac{H}{R}\right)} \;\; {\rm cm}^{-3},
\end{equation}
where $\mu$ is the mean mass per particle ($\sim 2.3$ in the molecular phase for cosmic abundances) in units of the mass of hydrogen $m_{\rm H}$. This is in the right density range (i.e. $\approx 10^8 - 10^{10}$ cm$^{-3}$) provided the aspect ratio of the disc satisfies $3 \times 10^{-3} \lesssim \frac{H}{R} \lesssim 0.3$. Further,  for such densities and for temperatures appropriate for molecules to survive (i.e. $\lesssim 3000$ K), support by radiation pressure must be significant, even dominant (whatever the solution). This conclusion is akin to the analysis by Pier \& Krolik (1992). Assuming a vertically self-gravitating disc in hydrostatic equilibrium, the mid-plane temperature is
\begin{equation}
T_{\rm mid} \sim \left(\frac{3 \pi G c \Sigma^2}{4 \sigma}\right)^{1/4} \simeq 2000 \;\; {\rm K},
\end{equation}
where $\sigma$ is the Stephan constant and $c$ is the speed of light. For a typical grain opacity of $1$ cm$^2$/g, the vertical optical depth is $\tau \gtrsim 200$, implying a temperature at the disc surface of $\sim T_{\rm mid}/\tau^{1/4} \simeq 600$ K, as required (Neufeld, Maloney \& Conger, 1994).

The second remark concerns the gravitational stability of the disc. For both solutions, the Toomre $Q$-parameter is of the order of unity. Actually, again at $0.65$ pc, we have
\begin{equation}
Q \sim \frac{\kappa c_{\rm s}}{\pi G \Sigma} \approx \frac{v_\phi T_{\rm mid}^2}{\pi G R} \sqrt{\frac{8\sigma H}{3c \Sigma^3}} \sim 2,
\end{equation}
where $\kappa \approx \Omega$ is the epicyclic frequency and $c_{\rm s}$ is the sound speed. It means that the outer disc is in a marginally stable state, and does not resemble a standard disc (Kumar, 1999). This property is predicted by the standard model at the parsec scale, whatever the accretion rate in the inner disc. Note also that the fact that our solutions have $\Sigma \propto R^{-1}$ and $v^{\rm disc}_\phi \sim cst$ (see Fig. \ref{fig:cmp}) is fully consistent with what we expect in an (radially) isothermal $Q-cst$ disc (e.g. Mestel, 1963).

\section{Concluding remarks}

We have reminded that a non-keplerian motion can arise from the combination of the gravitational attraction of a black hole and a disc. We propose that the ``sub-keplerian'' profile traced by water masers in NGC 1068 results from such an effect, a possibility outlined already in Greenhill et al. (1996). We have reported two simple physical solutions producing approximately the observed mean velocity exponent, with, accordingly, the parameters for the black hole and for the outer disc in this active galaxy. As outlined, these solutions are not unique; for instance, it is possible to fit the feature near 12 mas (see Fig. \ref{fig:cmp}), but $\Sigma$ is no more a power law. From this point of view, it would be worthwhile to explore the whole parameter space.

The analysis discussed here can be applied to other AGN. Note however that the system considered here is very simple and would need refinements. In particular, there is no physical model for the disc and the present approach works with the most minimal conditions. For instance, Eq.(\ref{eq:vphi}) should be supplemented with non-gravitational terms (pressure gradient, advection, viscosity, etc.). Further, the gravitational  attraction, possibly significant, of other components of the AGN (molecular torus, stellar cluster, galactic bulge, etc.) is not taken into account. Significant deviations to axi-symmetry (like a warp) could also change our results, quantitatively, and would make the problem much more tricky. With these restrictions, we have shown that it is feasible to ``invert'' the rotation curve in order to constrain the mass distribution in the central engine at the parsec scale. With more sophisticated or self-consistent investigations (e.g. Lodato \& Bertin, 2002), one should be able to connect these outer solutions to the innermost ones where accretion theories are very uncertain.

\begin{acknowledgements}
I am grateful to S. Collin, C. Boisson, F. Le Petit, and G. Lodato for stimulating discussions. I acknowledge James M. Moran, as the referee, for valuable comments and suggestions. 
\end{acknowledgements}

\end{document}